\documentclass[pre,  amsmath, amssymb, superscriptaddress]{revtex4}

\usepackage{graphicx}
\usepackage{dcolumn}
\usepackage{bm}
\usepackage{braket}
\usepackage{xcolor}

\usepackage{amsmath}

\usepackage{amsmath}
\usepackage{SIunits}

\newcommand{\beq}{\begin{equation}}
\newcommand{\eeq}{\end{equation}}

\begin{document}

\title{Osmolyte-Induced Protein Stability Changes Explained by Graph Theory}

\author{Mattia Miotto\footnote{\label{co} These authors contributed equally to the present work.} 
}

\affiliation{Center for Life Nano \& Neuro Science, Istituto Italiano di Tecnologia, Viale Regina Elena 291,  00161, Rome, Italy}

\author{Nina Warner \footnotemark[1] }
\affiliation{Yusuf Hamied Department of Chemistry, University of Cambridge}

\author{Giancarlo Ruocco}
\affiliation{Center for Life Nano \& Neuro Science, Istituto Italiano di Tecnologia, Viale Regina Elena 291,  00161, Rome, Italy}
\affiliation{Department of Physics, Sapienza University, Piazzale Aldo Moro 5, 00185, Rome, Italy}

\author{Gian Gaetano Tartaglia}
\affiliation{Department of Biology, Sapienza University of Rome, Rome, 00185, Italy.}
\affiliation{Center for Life Nano \& Neuro Science, Istituto Italiano di Tecnologia, Viale Regina Elena 291,  00161, Rome, Italy}

\author{Oren A. Scherman \footnote{\label{co2} These authors contributed equally to the present work.}
\footnote{\label{cr} Corresponding authors: \\oas23@cam.ac.uk\\ edoardo.milanetti@uniroma1.it}
}
\affiliation{Yusuf Hamied Department of Chemistry, University of Cambridge}

\author{Edoardo Milanetti \footnotemark[2] \footnotemark[3]}
\affiliation{Department of Physics, Sapienza University, Piazzale Aldo Moro 5, 00185, Rome, Italy}

\affiliation{Center for Life Nano \& Neuro Science, Istituto Italiano di Tecnologia, Viale Regina Elena 291,  00161, Rome, Italy}

\begin{abstract}
Enhanced stabilisation of protein structures via the presence of inert excipients is a key mechanism adopted both by  physiological systems and in biotechnological applications. %
While the intrinsic stability of proteins is ultimately fixed by their amino acid composition and organisation, the interactions between excipients and proteins together with their concentrations introduce an additional layer of complexity and in turn, method of modulating protein stability.
Here, we combined experimental measurements with molecular dynamics simulations and graph-theory based analyses to assess the stabilising/destabilising effects of different kinds of osmolytes on proteins during heat-mediated denaturation. 
We found that (i) proteins in solution with stability-enhancing osmolytes tend to have more compact interaction networks than those assumed in presence of destabilising excipients; (ii) a strong negative correlation (R = -0.85) characterises the relationship between the melting temperature $Tm$ and the preferential interaction coefficient defined by the radial distribution functions of osmolytes and water around the protein and (iii) a positive correlation exists between osmolyte-osmolyte clustering and the extent of preferential exclusion from the local domain of the protein, suggesting that exclusion may be driven by enhanced steric hindrance of aggregated osmolytes.
\end{abstract}

\maketitle

\subsection{Introduction}
Rigorous understanding of excipient-induced protein stability changes is a long-standing aim across both applied and fundamental biology. \cite{canchi_cosolvent_2013, kamerzell_proteinexcipient_2011} From an applied perspective, understanding the stabilising potential of therapeutically inert additives is key to developing sophisticated formulations for labile biologics; from fundamental standpoint, these interactions offer insight into the remarkable ability of life itself to persist in extreme environments by osmolyte-related mechanisms. \cite{lushchekina_impact_2020,rydeen_osmolytes_2018, khan_naturally_2010,lin_why_1994}

Efforts to explain osmolyte-induced protein stability changes have centred around several inter-related mechanisms which can be described generally by the preferential exclusion theory. \cite{ohtake_interactions_2011, ferreira_analyzing_2015} This theory, originally proposed by Serge Timasheff in  the 1980s, attributes osmolyte-derived stability to the preference for these molecules—-- largely on account of an unfavourable interaction with the peptide backbone--- to be excluded from the protein surface. \cite{gekko_mechanism_1981, arakawa_stabilization_1985, timasheff_water_1992,timasheff_control_1998, timasheff_protein_2002,timasheff_protein-solvent_2002} The thermodynamic drive to minimise solvent-exposed surface area results both in compaction of the native state and enhancement of the energetic barrier to unfolding, manifesting as an increase in protein denaturation temperature, Tm. On the contrary, destabilising osmolytes such as urea are believed to behave in an opposite manner, inducing a more diffuse structure and lower barrier to unfolding (suppressed Tm) via preferential interaction with the protein. \cite{canchi_cosolvent_2013}

Despite general acceptance, much of the molecular detail pertaining to the preferential exclusion theory remains poorly understood or hotly debated. \cite{canchi_cosolvent_2013} For example, the nature of the primary repulsive forces between the protein and solute remains contentious as does the role of osmolyte-mediated perturbations to the solvent network and concomitant perturbations to solvent-protein interactions. \cite{bolen_structure_2008} Modern computational techniques have helped to answer some of these questions, whilst occasionally raising more. Molecular dynamics (MD) simulations have produced mounting evidence attributing  urea-induced denaturation to preferential interaction with the protein, with variable degrees of importance given to secondary mechanisms involving indirect interactions through structural perturbations of the solvent. \cite{eberini_simulation_2011, canchi_cosolvent_2013, soper_impact_2003, lee_trehalose_2015, kuffel_hydrogen_2010, smolin_tmao_2017} Simulations have also shed light on the driving forces behind these interactions, providing evidence for steric and enthalpically motivated preferential exclusion and interaction, respectively. \cite{adamczak_effect_2018} Moreover, molecular understanding of osmolyte-induced stability changes has been greatly enriched through MD-based study of water destructuring, hydrogen bonding network perturbations (water-peptide backbone, water-side chain), water-protein/water-water/osmolyte-protein radial distribution functions (RDFs), protein hydration fraction, residue mean square fluctuations (RMSFs), and  protein/solvent relaxation times among other parameters. \cite{lushchekina_impact_2020, pazhang_effect_2016, lerbret_how_2007, mehrnejad_effects_2011, lins_trehaloseprotein_2004, liu_molecular_2010}

\begin{figure}[]
    \centering
    \includegraphics[width = \textwidth]{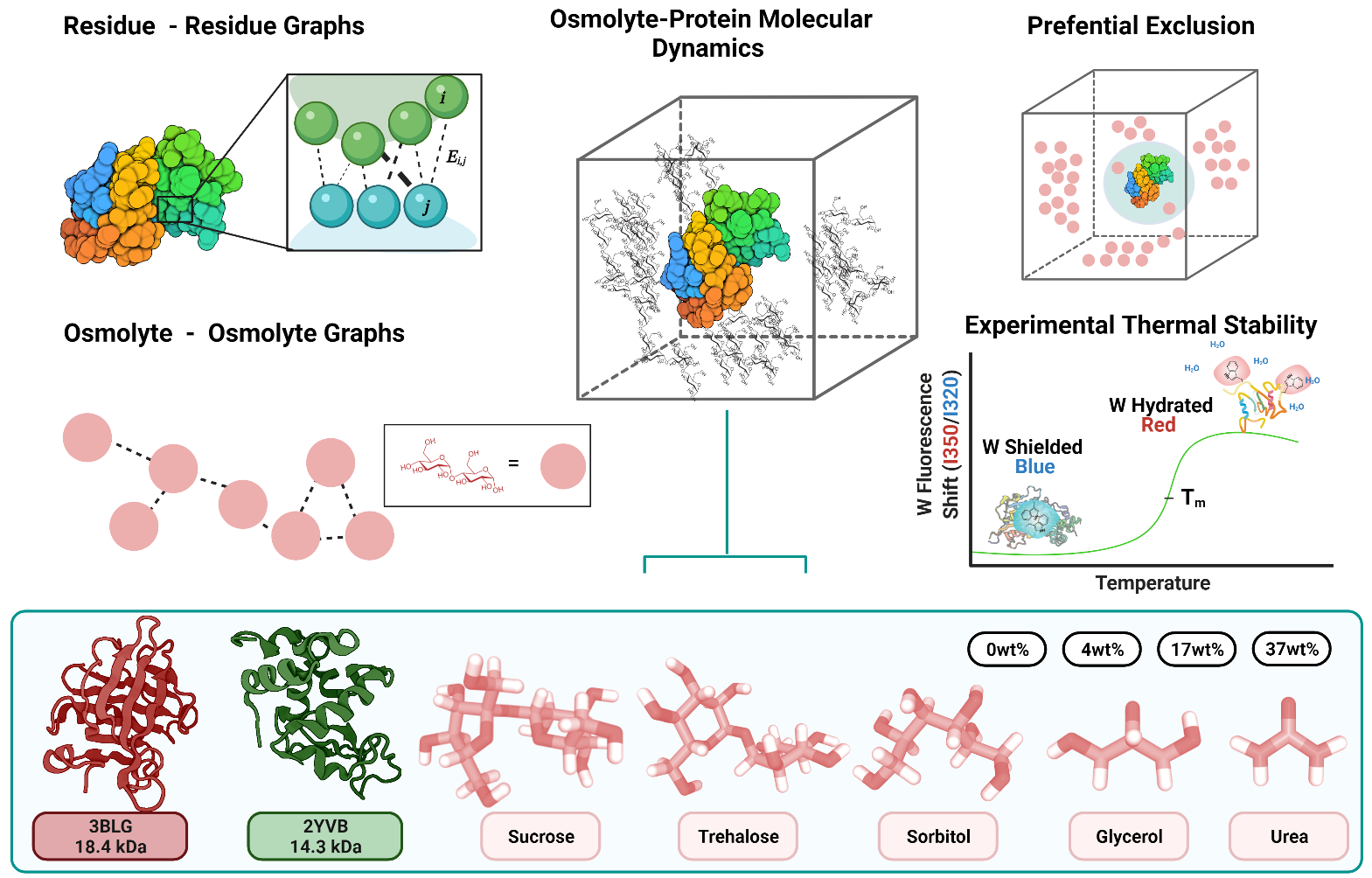}
    \caption[Scope of study.]{\textbf{Study overview and scope.} MD simulations were performed for thirty-two systems in total; two model proteins $\beta$-Lactoglobulin (PDB: 3BLG) and Lysozyme (PDB: 2YVB) were studied in aqueous solutions of four common protecting osmolytes: sucrose, trehalose, sorbitol, glycerol and one denaturing osmolyte: urea. All species were simulated at four concentrations: 0, 4, 17, and 37wt\%. Protein thermodynamic stability changes in the presence of osmolytes were measured experimentally by intrinsic tryptophan fluorescence (ITF) shifting. Simulations were analysed from a graph-theoretical lens with respect to both protein residue-residue interactions and osmolyte-osmolyte interactions as well as by residue dynamics, and osmolyte preferential exclusion coefficients. For more information, see Methods.}
    \label{RIN-overview}
\end{figure}

\begin{figure}[]
    \centering
    \includegraphics[width = \textwidth]{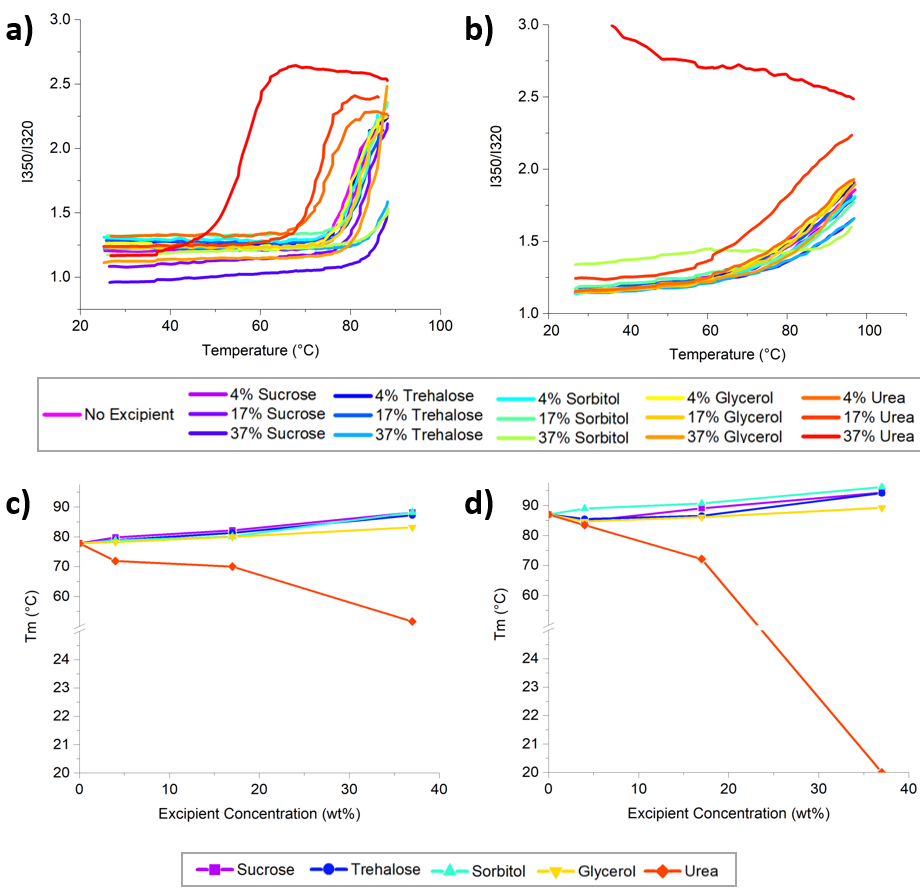}
    \caption{ \textbf{Experimental stability data for LYS  and $\beta$L  in the presence of osmolytes.} \textbf{a)} LYS ITF unfolding curves. \textbf{b)} $\beta$L ITF unfolding curves. \textbf{c)} Effect of osmolyte concentration on LYS Tm. \textbf{d)} Effect of osmolyte concentration on $\beta$L Tm. See Methods section for details on measurement and Tm calculation.}
    \label{Tm-lys+blact}
\end{figure}

Interestingly, many of these studies have observed self-aggregated clustering behaviour amongst stabilising osmolytes. \cite{pazhang_effect_2016, mehrnejad_effects_2011, liu_molecular_2010, lins_trehaloseprotein_2004} These clusters have been studied in the absence of protein on various occasions. Lee et al. showed that applying spectral graph analysis to three binary water-osmolyte simulations (water-urea, water-sorbitol, and water-trimethylglycine) could reveal distinct morphological differences between the graph networks formed by the protecting (sorbitol and TMG) and denaturing (urea) osmolyte classes. In the former case, the authors observed the formation of extended networks, descriptively similar to that of water, whereas even at high concentrations, urea failed to form a continuous network and instead assembled only into small, segregated clusters. \cite{lee_spectral_2015} Similarly, persistent homology based topological characterisation of urea and TMAO aggregates by Mu and colleagues revealed morphological differences in aggregates formed by the chaotropic and protective osmolyte species. \cite{xia_persistent_2019,anand_weighted_2020} A recent report by Sundar et al. presented an exhaustive graphical analysis of seven aqueous osmolyte solutions and found no significant perturbations to the water network when properly accounting for the contribution of water-osmolyte interactions. From their analysis, the authors attributed osmolyte-induced protein stability changes to direct interaction between the protein and osmolyte. Unfortunately, however, the absence of protein in the simulated solutions precluded direct assessment of this hypothesis. \cite{sundar_unraveling_2021}

\begin{figure}[]
    \centering
\includegraphics[width = \textwidth]{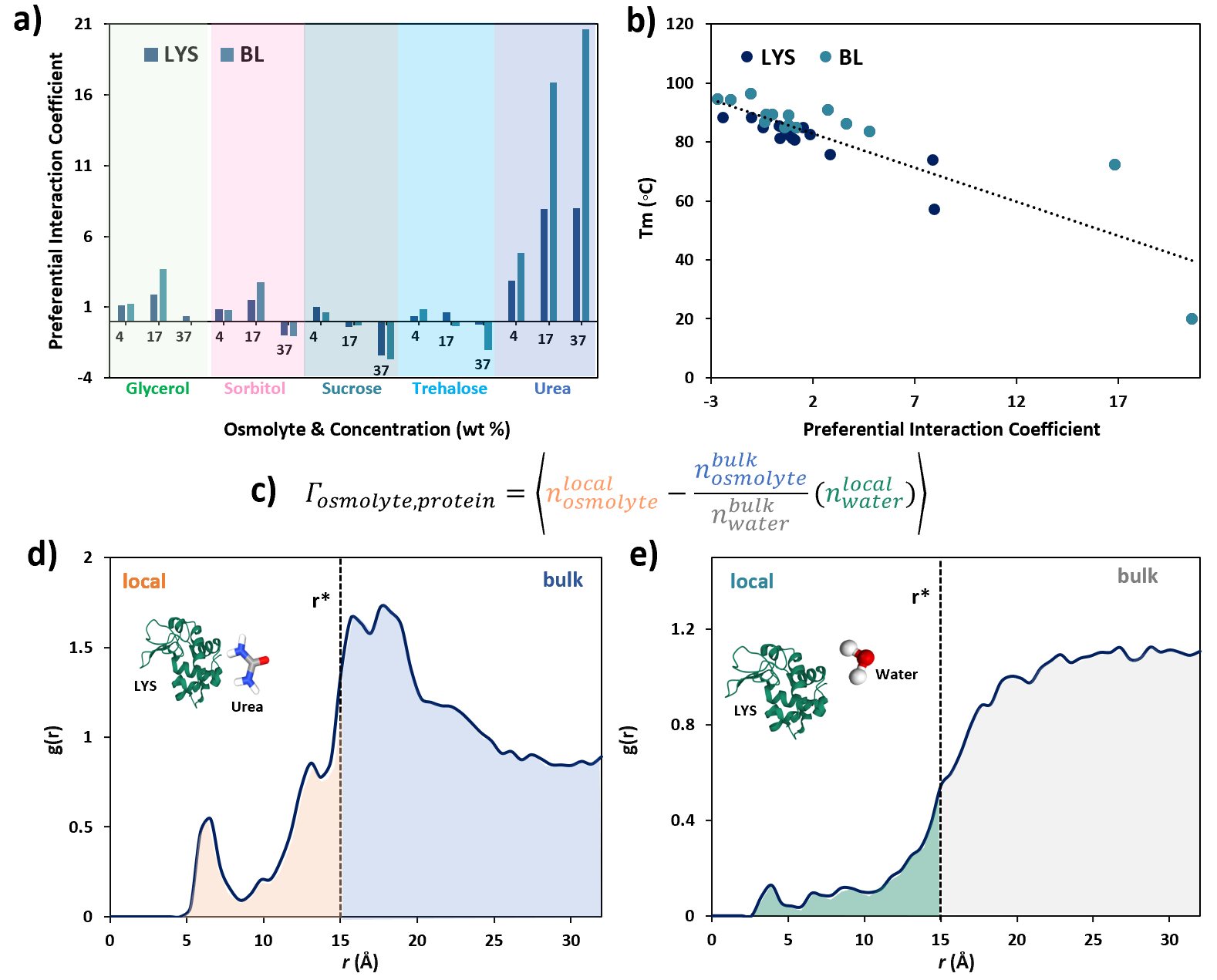}
 \caption{ \textbf{Analysis on osmolyte-protein preferential interaction.} \textbf{a)} Osmolyte-protein preferential interaction coefficients ($\Gamma_{o,p}$) for all 30 simulated protein-osmolyte systems. \textbf{b)} $\Gamma_{o,p}$ v. Tm. $R = -0.85$. P-value (points): 0.00001 (24). \textbf{c)} Equation used to calculate $\Gamma_{o,p}$. n is number of molecules belonging to a particular class (osmolyte/water, subscript) in a specified domain (local/bulk, superscript). The calculation of n is achieved by integration of g(r). \textbf{(d)} and \textbf{(e)} depict the protein-osmolyte and protein-water g(r) of an exemplary system (LYS in 37 wt\% urea) respectively. Shaded regions below the g(r) curves are coloured to match their corresponding terms in the $\Gamma_{o,p}$ expression. The local domain is defined by r = [0,r*] and the bulk domain by r = [r*, r$_{box\:boundary}$]. The cutoff distance between the local and bulk domains, r*, is selected as described in the Methods.}
\label{prefexclusion}
\end{figure}

Over the past two decades, graph representations of proteins have proved powerful models for unearthing molecular origins of stability. Analyses of residue interaction networks (RINs)--- or energy-weighted networks of non-covalent interactions between non-adjacent amino acids (nodes)--- have shown good success in identifying stability-linked residues, structural elements, and network descriptors. \cite{vishveshwara_protein_2002,brinda_network_2005, fernandez_classification_2008,giollo_neemo_2014, pires_mcsm_2014, Miotto2020_epidemics} In addition to providing mechanistic insight into stability, RINs have been used extensively in a predictive context, particularly relating to the effect of mutations on protein stability; indeed, the first example of ab initio structure-based stability predictions were reported in 2019 using an RIN framework. \cite{miotto_insights_2019, Miotto2022_server, Desantis2022}

Encouraged by both the success of RINs in elucidating protein stability at a molecular level and the distinct morphological differences in graphical networks of protecting and denaturing osmolytes, a series of protein-osmolyte systems were investigated from a graph-theoretical perspective. \cite{lee_spectral_2015} 
As far as the authors are aware, this is the first graph theoretical study of protein-osmolyte solutions. 
Herein, the thermal stability of two model proteins in a series of industrially and biologically relevant osmolyte solutions is described.
Each system is probed by a combined experimental-computational approach encompassing intrinsic tryptophan fluorescence, molecular dynamics (MD) simulation, and graph analysis of both protein residue interactions and osmolyte interactions.
These results are discussed in reference to modern osmolyte theory; a graph-theoretical descriptor is introduced to predict osmolyte-induced stability changes.
Finally, the clustering behaviour of osmolytes is related to the seminal theory of preferential exclusion and a novel mechanism by which stabilising osmolytes are excluded from the protein surface is proposed.

\section{Results and Discussion}
To investigate the effects of osmolyte-induced protein thermal stability with generalisability, two model proteins were studied--- a mainly alpha species, Lysozyme (LYS, PDB:2YVB), and a  mainly beta species, $\beta$-Lactoglobulin ($\beta$L, PDB:3BLG). 
Both model proteins were relatively small (LYS: 14.3 kDa, BLG: 18.4 kDa) in the interest of computational efficiency. 
LYS and BLG were studied by experiment and MD simulation (100ns) in the presence of both protecting (sucrose, trehalose, sorbitol, glycerol) and denaturing (urea) osmolyte solutions of 0, 4, 17, and 37 wt\%.
An overview of the study discussed herein is provided in Figure \ref{RIN-overview}.

\subsection{Characterisation of Protein Thermal Stability \label{os-thermalstability}}

Osmolyte-induced thermodynamic protein stability changes were studied by ITF. [Figure \ref{Tm-lys+blact}] 
As expected, as the concentrations of protecting osmolytes were increased, protein Tm followed; addition of urea, in contrast, induced deleterious effects on the thermal stability of both proteins. [Figure \ref{Tm-lys+blact}b and d] 
This was more pronounced in the case of $\beta$L, for which high urea concentrations (37wt\%) induced denaturation in the absence of heating. [Figure \ref{Tm-lys+blact}c] 
For this sample, Tm was approximated as 20 ${}^o$C (room temperature). [Figure \ref{Tm-lys+blact}d]

The extent of concentration-normalised protection conferred by osmolytes studied is depicted in Figures \ref{Tm-lys+blact} b and d.
Interestingly, the relative efficacy of protecting osmolytes in raising protein Tm was found to depend on concentration. In the case of LYS, sucrose conferred the best stabilisation at 4 and 17 wt\%, however at the highest concentration studied (37wt\%), the effects of sucrose, sorbitol, and trehalose on LYS Tm fell within error ($\Delta_{Tm}$ = +/- 1${}^o$ C) of one another. 
In contrast, in the case of $\beta$L, sorbitol conferred the best stability across all three concentrations. 
For both $\beta$L and LYS, glycerol exhibited the poorest stabilisation of all protecting osmolytes studied. 
Furthermore, in both cases, variance in Tm amongst all protecting osmolyte systems was roughly 10${}^o$ C, indicating comparable magnitudes of stabilisation.

Interestingly, room temperature ITF measurements--- indicative of the extent of structural perturbation induced by the osmolyte in the absence of heat--- did not appear to predict Tm. 
This was particularly evident in formulations containing sorbitol, for which relatively high red shifting at room temperature was observed together with high Tm.
This may suggest that sorbitol stabilises proteins in a manner that is mechanistically distinct from the other osmolytes.

\subsection{Protein-Osmolyte g(r)}

\begin{figure}[]
    \centering
    \includegraphics[width = \textwidth]{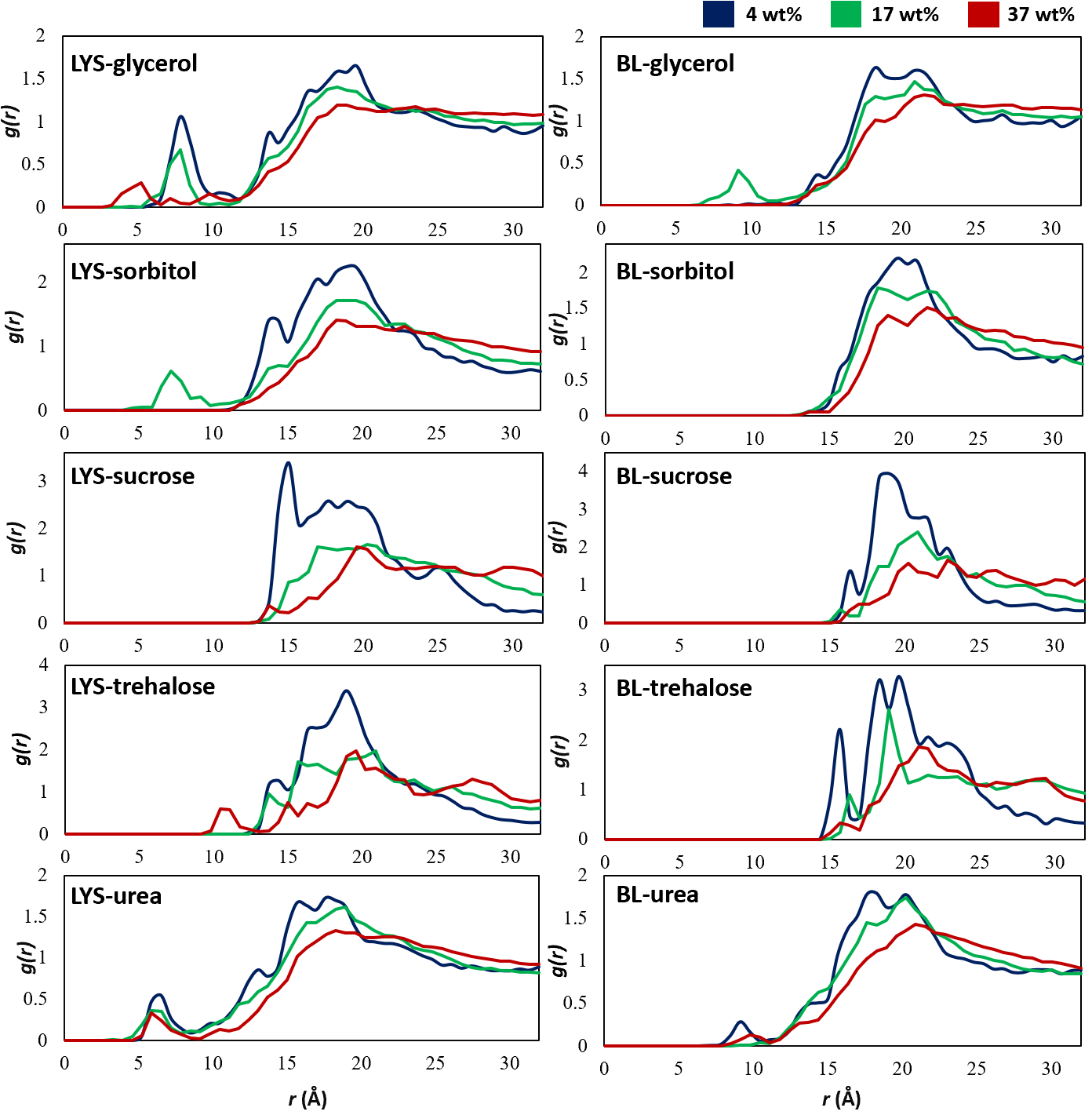}
    \caption{Osmolyte-protein radial distribution functions (g(r)). Line colour indicates osmolyte concentration (navy: 4 wt\%, green: 17 wt\%, red: 37wt\%). See Methods for details on radial distribution function (g(r)) calculation.}
    \label{RDF_protein_os}
\end{figure}

Concentration-dependent protein-osmolyte interactions were studied by MD.
Figure \ref{RDF_protein_os} 
depicts the radial distribution functions (g(r))--- the relative density of osmolyte molecules as a function of distance from the protein--- for all systems studied.
Comparative inspection of these graphs reveals several key distinctions amongst osmolyte-protein interactions.
First, in general, it is evident that glycerol and urea present well defined peaks with high proximity to the protein surface relative to other osmolytes; the intensity and position of these peaks, however, appear to be protein-dependent.
Furthermore, poor conservation of the shape of g(r) across various concentrations of a single osmolyte suggests a concentration dependence of protein-osmolyte interactions; this is least apparent for urea, in the case of which minimal peak shifting or emergence is observed.
Finally, it is interesting to note that in disaccharide (trehalose, sucrose) systems, several well-defined peaks are evident at a distance $\geq$ 15{\AA} from the protein.
These suggest the presence of ordered osmolyte structures well beyond the first hydration layer.

\subsection{Preferential Exclusion of Osmolytes}
Protein-osmolyte preferential interaction coefficients were calculated from the protein-osmolyte and protein-water radial distribution functions as described in the Methods Section. Positive preferential interaction coefficients indicate an excess of the osmolyte in the local domain of the protein, whilst a negative preferential interaction coefficient indicates exclusion of the osmolyte from the direct surroundings of the protein (and a relative preference for the bulk phase). Preferential interaction coefficients for all protein-osmolyte systems are presented in Figure \ref{prefexclusion}.

\begin{figure}[]
    \centering
\includegraphics[width= \textwidth]{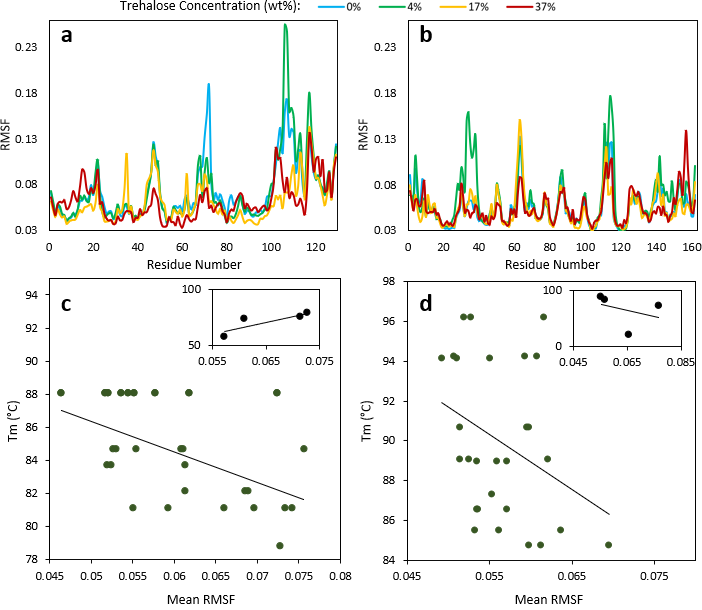}
    \caption[Mean RMSF v. Tm]{\textbf{Mean RMSF v. Tm for proteins in the presence of osmolytes.} Example of raw RMSF data at various trehalose concentrations for \textbf{a)} LYS and \textbf{b)} $\beta$L. Mean RMSF v. Tm  considering protecting and destabilising (inset) osmolytes for \textbf{c)} LYS and \textbf{d)} $\beta$L.  
    The correlation coefficient relating mean RMSF and Tm for LYS in the presence of protecting osmolytes (\textbf{c)} is R = -0.52 (p-value: 0.005, points: 27); in the presence of destabilising osmolyte (urea, inset), R = +0.83 (p-value: 0.17, points: 4).
    The correlation coefficient relating mean RMSF and Tm for $\beta$L in the presence of protecting osmolytes (\textbf{d)} is R = -0.34 (p-value: 0.08, points: 27); in the presence of destabilising osmolyte (urea, inset), R = -0.34 (p-value: 0.66, points: 4).}
    \label{RMSF-Tm}
\end{figure}

Consistent with previous findings \cite{shukla_molecular_2009, canchi_cosolvent_2013}, urea presented a positive preferential interaction coefficient with both proteins at all concentrations. [Figure \ref{prefexclusion}]
The preferential interaction coefficient characterising urea/$\beta$L was higher than that characterising urea/LYS across all concentrations; this difference was most dramatic at 37wt\%, however, reflecting the relative magnitude of Tm depression observed amongst the two systems ($\beta$L: $\Delta_{Tm}$ = - 65, LYS: $\Delta_{Tm}$ = -26).
The origin of the differential extent of preferential interaction amongst urea/$\beta$L and urea/LYS  (as well as differences amongst coefficients characterising interactions between the two proteins and protecting osmolytes) was not the primary focus of this study and was thus not investigated. 
Nonetheless, a systematic investigation into the relationship between protein size, degree of backbone solvation, surface residue composition, or three-dimensional shape/topology and preferential interaction coefficients could be a fascinating area of future research.

Amongst all protecting osmolytes studied, the degree of exclusion increased with concentration.
Generally, sucrose and trehalose were most strongly excluded from the local environment of the protein; interestingly, the protein-osmolyte radial distribution functions of these species also showed the greatest degree of structural organisation at distances $\geq$ 15 {\AA} from the protein (Figure ).
%
Such an observation might suggest exclusion of relatively ordered disaccharide assemblies; this hypothesis is investigated in detail in Section \ref{pref-os-clusters}.
It is also noted that glycerol--- the least effect protecting osmolyte by concentration-dependent melting temperature elevation--- is the only protecting species not preferentially excluded at high concentration (37 wt\%). [Figure \ref{prefexclusion}]

To more quantitatively assess the relationship between osmolyte-induced thermal stability changes and preferential exclusion, the preferential interaction coefficients of all osmolyte-protein combinations were plotted against protein melting temperatures (Tm).
A strong negative correlation (R = -0.85) was found to characterise the relationship between Tm and preferential interaction coefficient, consistent with preferential exclusion theory. \cite{timasheff_protein-solvent_2002, gekko_mechanism_1981, canchi_cosolvent_2013}

\begin{figure}[]
    \centering
\includegraphics[width= 0.70\textwidth]{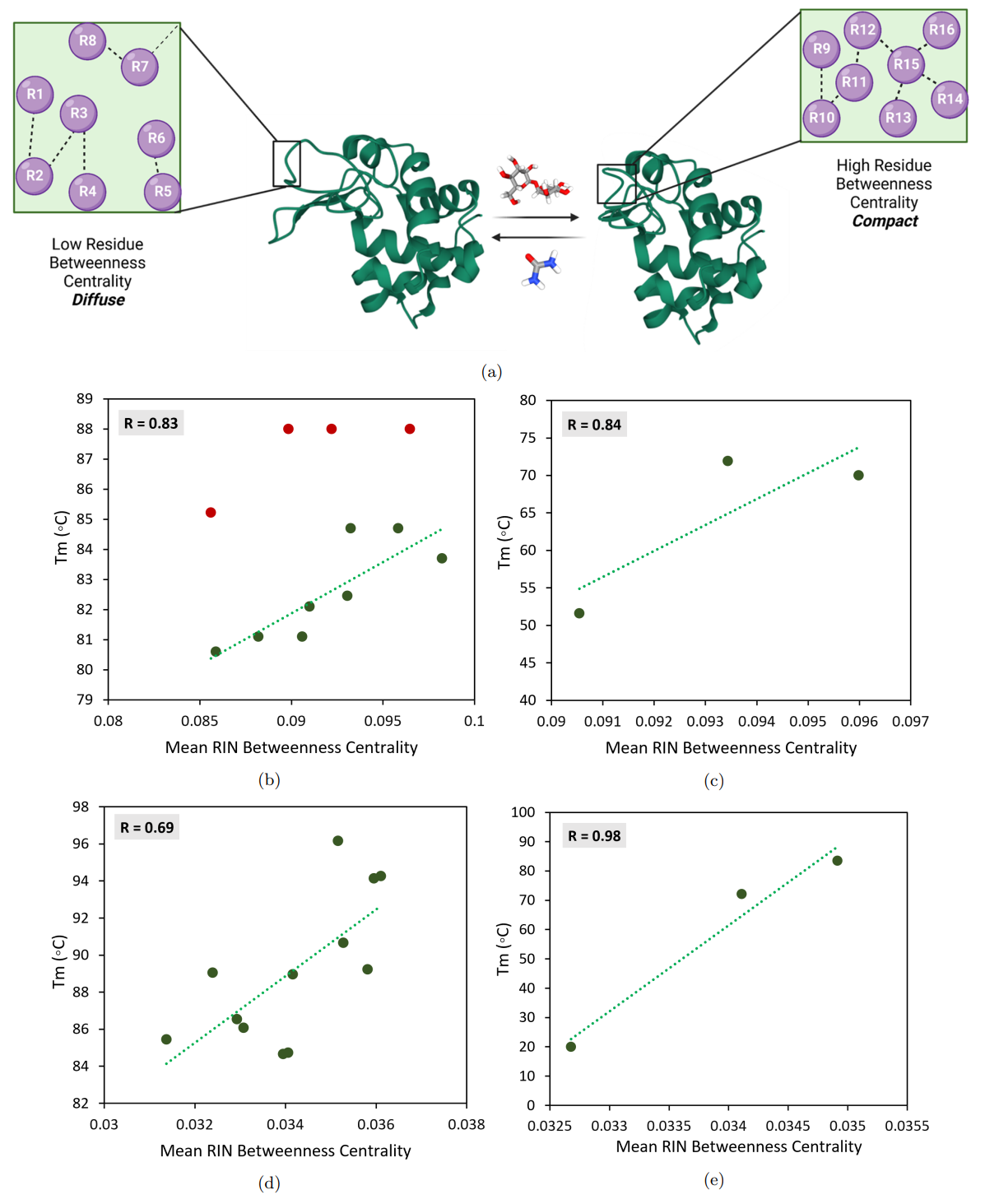}
\caption[Mean RIN Betweenness Centrality v. Tm]{\textbf{Mean RIN BC v. Tm for proteins in the presence of protecting and denaturing osmolytes.} \textbf{a)} Illustration of RIN BC for protein stability monitoring \textbf{b)} Mean RIN BC v. Tm for LYS in solutions of protecting osmolytes and \textbf{c)} urea. \textbf{d)} Mean RIN BC v. Tm for $\beta$L in solutions of protecting osmolytes and \textbf{e)} urea. Mean RIN BC, like RMSF, poorly predicted Tm at high concentrations of protecting osmolytes for LYS; however, this was not the case for $\beta$L, for which a moderate correlation (R = 0.69) between RIN BC and Tm was evident across all concentrations. Data collected on LYS in the most concentrated osmolyte solutions is depicted in red \textbf{b)}; these datapoints were excluded when calculating the reported correlation coefficient (R = 0.83). The correlation coefficient between RIN BC and Tm for LYS across all protecting osmolyte solutions was 0.33.
P-values (points): \textbf{b)} 0.011 (8), \textbf{c)} 0.36 (3), \textbf{d)} 0.013 (12), \textbf{e)} 0.13 (3). \label{BC-RIN-Tm}}
\end{figure}

Correlations between osmolyte-protein preferential interaction coefficients and Tm were further disaggregated by osmolyte/protein pair; Table \ref{pref_ex_tm_table} reports R values for the relationships between Tm and preferential interaction coefficients across all ten protein/osmolyte combinations.

\begin{figure}[]
    \centering
\includegraphics[width= 0.80\textwidth]{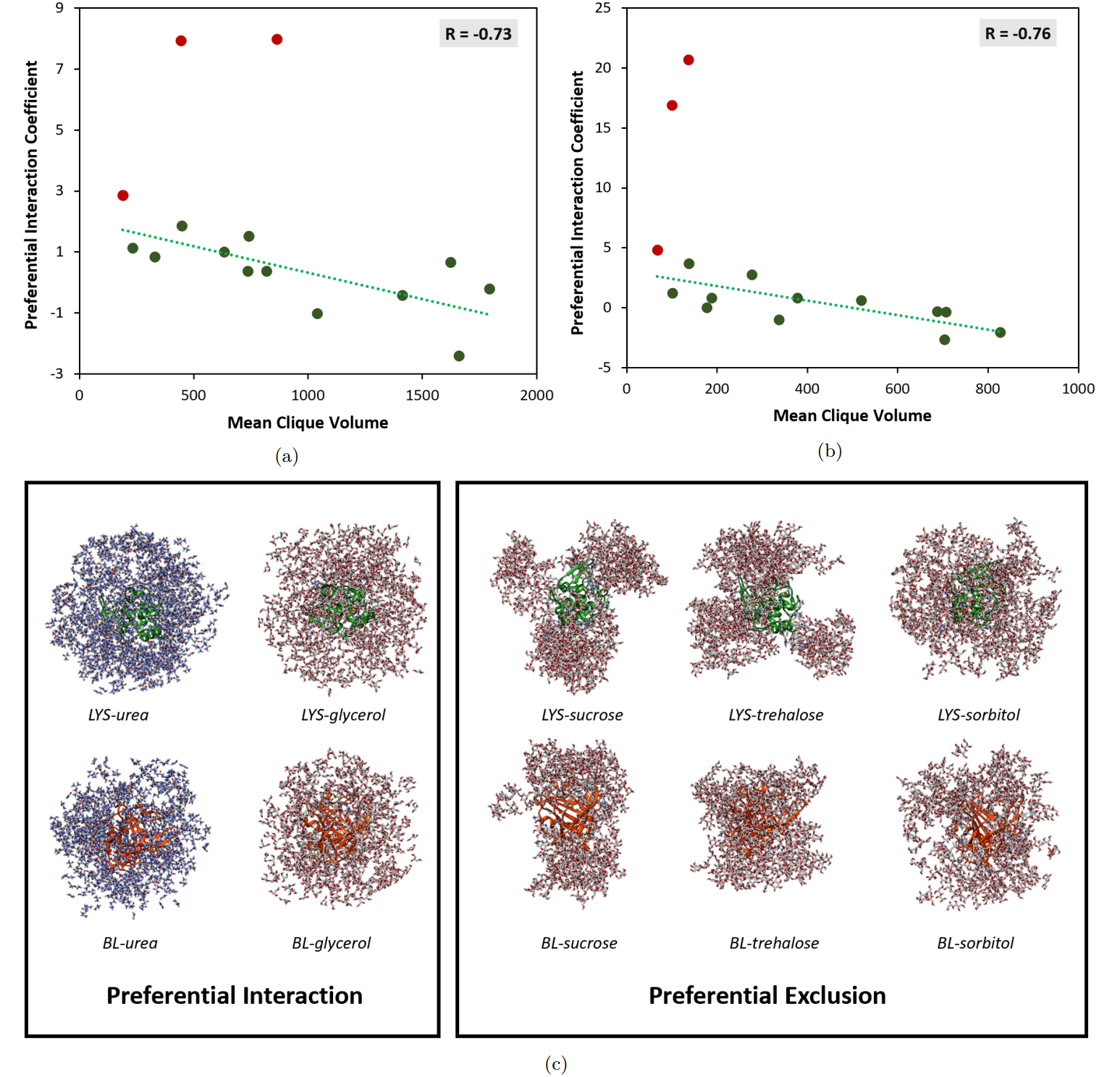}
\caption{\textbf{Osmolyte clique volume v. preferential interaction coefficient.} \textbf{a)} Mean osmolyte clique volume v. preferential interaction coefficient for LYS. \textbf{b)} Same as in a) but for$\beta$L. All concentrations (4, 17, 37 wt\%) of protecting osmolytes (glycerol, sorbitol, sucrose, trehalose) and the lowest concentration (4 wt\%) of urea were included in the reported correlation coefficients (R = -0.68 and -0.72 for LYS and $\beta$L, respectively). The units of mean clique volume are {\AA}$^{3}$. P-values (points): \textbf{a)} 0.007 (12), \textbf{b)} 0.0041 (12). \textbf{c)} MD snapshots of osmolytes with positive (glycerol, urea) and negative (sorbitol, sucrose, trehalose) preferential interaction coefficients at 37 wt\%. Preferentially excluded osmolytes tend to form osmolyte-osmolyte clusters whilst preferentially interacting species lack ordered domains within the bulk phase, appearing to exhibit no preference for one region of the bulk domain over another. Furthermore, unlinked osmolytes (unimers and dimers) are visually apparent in preferentially interacting systems, while preferentially excluded osmolytes appear as subunits of extended networks.
\label{mcv_pi}}
\end{figure}

In all cases, the relationship between Tm and preferential interaction was consistent (negative R value). [Table \ref{pref_ex_tm_table}]
Interestingly, the osmolyte-protein pairs for which the largest (by absolute value) preferential interaction coefficients were calculated ($\beta$L/urea, $\beta$L/trehalose, $\beta$L/sucrose and LYS/sucrose) were also characterised by the strongest correlations between Tm and preferential interaction.
This suggests that when a significant difference in osmolyte preference for the local (protein) and bulk domains is absent, secondary factors likely become the predominant predictors of Tm.

\begin{table}[]
\centering
\caption{Tm v. Preferential Interaction Correlation Coefficients (R) by Osmolyte-Protein Pair.}
\label{pref_ex_tm_table}
\resizebox{0.2\textwidth}{!}{\begin{tabular}{ccc}
\hline
&LYS                                        & $\beta$L                                      \\ \hline
Glycerol             & -0.59                                      & -0.51                                         \\
Sorbitol             & -0.75                                      & -0.72                                         \\
Sucrose              & -0.99                                      & -0.98                                         \\
Trehalose            & -0.77                                      & -0.95                                         \\
Urea                 & -0.58                                      & -0.80                                         \\ \hline
\end{tabular}}
\end{table}

\subsection{Protein Dynamics in Osmolyte Solutions}

Neither $\beta$L nor LYS showed a significant change in structure (secondary or tertiary when modelled in the presence of osmolytes at room temperature. 
Surprisingly, this was even the case when proteins were simulated for a microsecond in 37 wt\% urea, despite $\beta$L being fully denatured under these conditions when studied by experiment. [Figure \ref{Tm-lys+blact}]
Similar observations of unperturbed LYS and $\beta$L native structure simulated in the presence of concentrated urea at room temperature have been reported by both Biwas et al. and Eberini et al.. \cite{biswas_contrasting_2018, eberini_simulation_2011}

Residue mean square fluctuation (RMSF) is a common metric for assessing relative stability of simulated proteins below the threshold of detectable secondary or tertiary structural change. \cite{dong_structural_2018}
We studied the relationship between protein mean RMSF and Tm to discover if room temperature protein dynamics-- unlike structural changes-- could predict the thermodynamic stability of proteins in the presence of osmolytes.
For systems containing protecting osmolytes, strong negative correlations (R = -0.74 and -0.73 for LYS and $\beta$L, respectively) were found to relate mean RMSF and Tm for systems at low osmolyte concentration (4 and 17wt\%). [Figure \ref{RMSF-Tm}a,c]
This relationship, however broke down at high concentrations of protecting osmolyte (37 wt\%); indeed, when datapoints corresponding to this concentration were included in the model (red data points in Figures \ref{RMSF-Tm}a and c), the correlations between mean RMSF and Tm for both proteins were considerably weakened (R = -0.45 and -0.21 for LYS and $\beta$L, respectively).

Furthermore, when studied in the context of urea, the relationship between RMSF and Tm failed to extrapolate.
In fact, while the correlation between mean RMSF and Tm for $\beta$L was not significant (R = -0.11), a near equal and opposite positive correlation was observed between protein dynamics and Tm for LYS/urea systems (R = 0.76). [Figure \ref{RMSF-Tm}b,d]
This may be related to bonding between LYS surface residues and urea, leading to reduced local flexibility of the protein.

Independent of origin, however, the inconsistent relationship between mean RMSF and Tm confounds the interpretation of RMSF in osmolyte systems with respect to protein stability.

\subsection{Osmolyte-Induced Perturbations to Protein Graphs}

Residue interaction networks (RINs) have recently been shown to predict protein thermostability \textit{ab initio}. \cite{miotto_insights_2019} Thus, we hypothesised that analysing the RINs of $\beta$L and LYS in the presence of different osmolytes may afford a superior method for probing stability than conventional methods (tertiary/secondary structure or mean RMSF analysis).

The networks of non-covalent interactions amongst protein residues were modelled as graphs wherein nodes were defined by the C$\alpha$ atoms of amino acids, covalent interactions were ignored, links between (non-adjacent) nodes existed if the distance between them was $\leq$ 9{\AA}, and the weight of each link was defined by either the van der Waals or Coulombic interaction energy. [See Methods Section for more details]

The RIN mean betweenness centrality (BC)--- a measure of the frequency with which a node appears amongst the shortest paths connecting all other possible node pairs in the network--- was found to be more predictive of osmolyte-induced protein stability changes (Tm) than either RMSF or conventional structure measures.
A strong, positive correlation was found to relate mean RIN BC and Tm, suggesting that osmolyte-induced protein stability enhancement can be largely attributed to increased BC amongst protein residues, a proxy for structural compaction.
Like RMSF, however, in the case of LYS, this trend was found to somewhat collapse at high osmolyte concentration; inclusion of datapoints corresponding to 37 wt\% osmolyte led to a dampening of the correlation coefficient from R = 0.83 to R = 0.33. [Figure \ref{BC-RIN-Tm}a]
Despite this, unlike RMSF, mean RIN BC demonstrated good predictability of Tm across all osmolyte concentrations for $\beta$L and extrapolated to denaturing systems (urea) for which a consistent strong positive correlation was observed for both proteins (R = 0.84 and R = 0.98 for LYS and $\beta$L, respectively). [Figure \ref{BC-RIN-Tm}c, d]
These results suggest that mean RIN BC offers an alternative descriptor for monitoring osmolyte-induced room temperature perturbations to protein thermodynamic stability often too subtle to detect by conventional metrics.

\begin{figure}[]
    \centering
    \includegraphics[width = \textwidth]{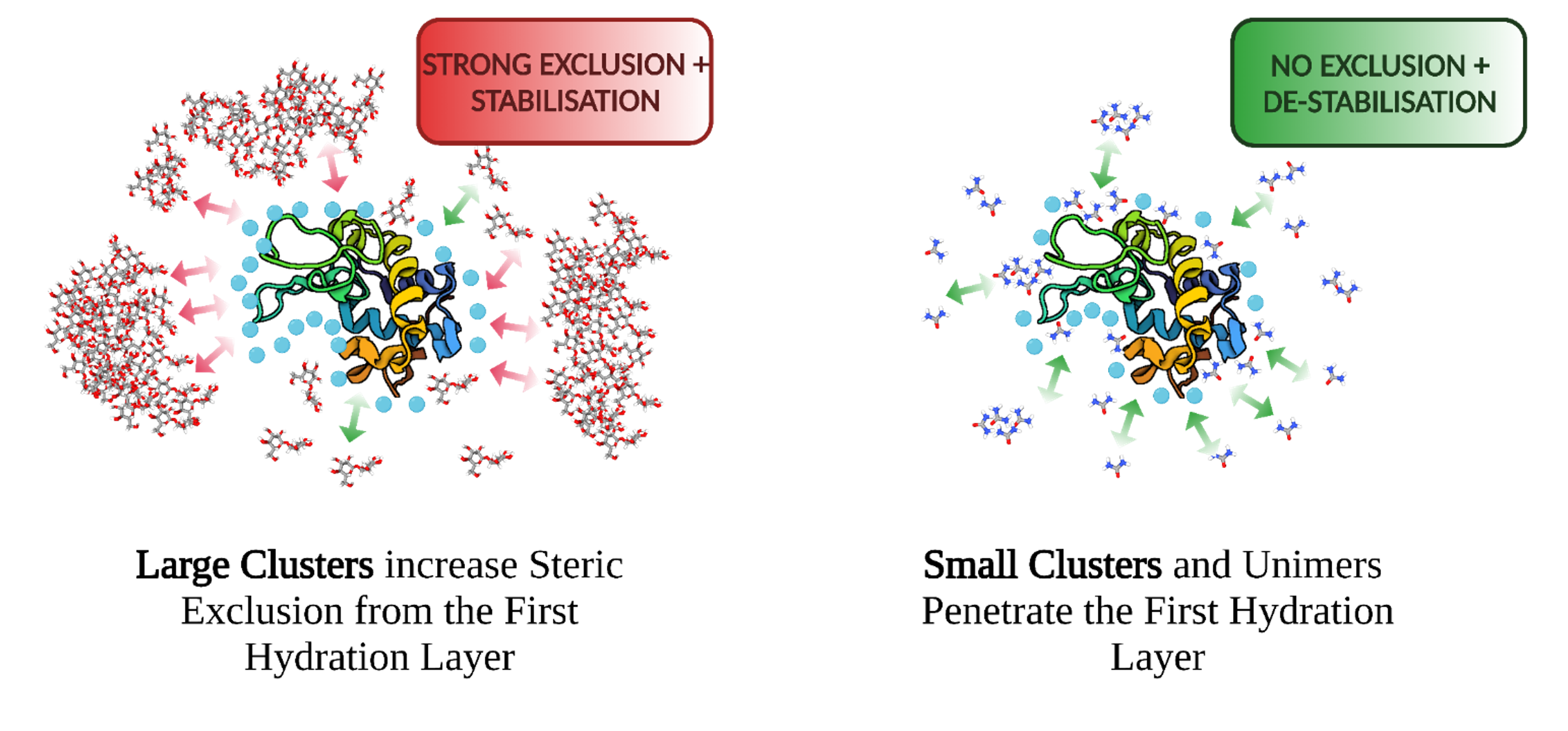}
    \caption{Hypothesis for cluster-driven preferential exclusion of osmolytes.}
    \label{cluster-hypothesis}
\end{figure}

\subsection{Preferential Exclusion of Osmolyte Clusters\label{pref-os-clusters}}

To better understand the relationship between osmolyte clustering and preferential exclusion, osmolyte-osmolyte networks were analysed as unweighted graphs since osmolyte are of the same kind in each considered system (See Methods). 
Interestingly, the mean clique volume (i.e. the volume of fully connected subgraphs or clusters) was found to be predictive of the osmolyte-protein preferential interaction coefficient. [Figure \ref{mcv_pi}]
This relationship was characterised by strong, negative correlation coefficients for both proteins (R = -0.73 and -0.76 for LYS and $\beta$L, respectively), suggesting that preferential exclusion may be driven to a significant extent by steric exclusion of osmolyte clusters--- rather than discrete molecules--- from the local domain of the protein.
Furthermore, the relationship between preferential interaction coefficients and mean osmolyte clique volume was consistent across all concentrations of protecting osmolytes, however failed to linearly extrapolate to systems containing high urea concentrations (17 and 37 wt\%).

We hypothesised that this inconsistency in trend observed at high urea concentrations could be accounted for by increasing node density, and in turn, node-to-node proximity. 
Thus, to disentangle the observed correlation from the effect of concentration (and concomitant effect of changes to global node density), the correlation between mean clique volume and preferential interaction was analysed across concentration normalised data. [Table \ref{table-cliquevol-pi}]
Indeed, a moderately strong, negative correlation (-0.61 $\leq$ R $\leq$ -0.71) was observed amongst all concentrations.

\begin{table}[]
\centering
\caption{Mean Clique Volume v. Preferential Interaction Coefficient Correlation (R) by Concentration}
\label{table-cliquevol-pi}
\resizebox{0.33\textwidth}{!}{%
\begin{tabular}{ccc}
\hline
\begin{tabular}[c]{@{}c@{}}Osmolyte \\ Concentration (wt\%)\end{tabular} & LYS   & $\beta$L \\ \hline
4                                                                        & -0.67 & -0.61    \\
17                                                                       & -0.71 & -0.73    \\
37                                                                       & -0.66 & -0.61    \\ \hline
\end{tabular}%
}
\end{table}

Additionally, the correlation between preferential exclusion and osmolyte clustering could be visually comprehended. 
Figure \ref{mcv_pi}c depicts MD snapshots of systems (37 wt\% osmolyte) exhibiting preferential interaction (left) and preferential exclusion (right).
In general, preferentially interacting osmolytes did not aggregate; instead, they appeared to diffuse across the entirety of simulation space with little preference for one region of the bulk domain over another.
In contrast, strongly excluded osmolytes (sucrose, trehalose) formed extended structural networks around the protein, failing to diffuse into the whole of simulation space.
Sorbitol, weakly excluded from both proteins, in contrast fell somewhere between these two morphological scenarios, reflective of the osmolyte's aberrant experimental behaviour discussed in Section \ref{os-thermalstability}.

Closer inspection of the snapshots in Figure \ref{mcv_pi}c reveals yet more granular distinctions between the two classes of osmolytes.
Examination of the osmolyte-protein system boundaries reveals the presence of isolated unimers and dimers in preferentially interacting species, whilst those characterised by preferential exclusion appear linked to a larger network.
Lastly, it is interesting to note the strong dependence of the morphology of aggregates formed by strongly excluded osmolytes (sucrose, trehalose) on the protein identity.
This could suggest that the protein itself governs osmolyte assembly, perhaps by indirectly acting as a scaffold via the first hydration layer.

\section{Conclusions and Outlook}
Two major conclusions can be drawn from this work; the first is that RIN BC is a high performing metric for monitoring subtle osmolyte-induced, room temperature structural perturbations by MD.
RIN BC shows a consistent, moderately strong negative correlation with experimental thermal stability (Tm) across systems containing protecting or denaturing osmolytes; in contrast, conventional methods such as mean RMSF and secondary/tertiary structure analysis fail to predict Tm with generalisability across both osmolyte classes.
Furthermore, alongside offering predictive utility, analysis of RIN BC offers insight into the mechanistic origins of osmolyte-induced protein stability--- structural compaction--- consistent with preferential exclusion theory. \cite{canchi_cosolvent_2013, timasheff_protein_2002}
Future investigations into the mechanistic relevance/predictive power of RIN BC in protein systems lacking osmolytes (i.e. proteins at high temperature or in the presence of other classes of stabilisers, etc.) would be necessary to elucidate the generalisability of this metric.

Second, it has been shown that a positive correlation exists between an osmolyte's clustering and the extent to which it is preferentially excluded from the local domain of the protein.
This correlation is further shown to be dependent upon cluster size, with clusters of larger volume being more strongly excluded from the protein.
These results suggest that the preferential exclusion phenomenon--- to which thermodynamic protein stabilisation by protecting osmolytes is attributed--- is likely, at least in part, motivated by the steric exclusion of molecular networks rather than unimeric species. [Figure \ref{cluster-hypothesis}]

Finally, it is noted that the morphology of clusters formed by excluded osmolytes appears to be protein-dependent.
This is intriguing, as--- by nature of being preferentially excluded--- these species do not significantly interact.
Thus, the relationship between the structure of osmolyte aggregates and protein identity is likely rooted in indirect relations, e.g. via the first hydration layer.
Such an explanation would suggest that the shape of the protein and in turn, the architecture of the primary hydration layer, governs the morphology of osmolyte clusters via a scaffolding effect.
An investigation into this hypothesis could serve as a fascinating premise for future study.

\section{Materials and Methods}

\subsection{Materials}

All materials were purchased from Sigma Aldrich unless otherwise noted. 

\subsection{Measurement of Protein Thermal Stability}
Protein conformational changes were monitored in duplicate by tryptophan fluorescence using a Cary Eclipse Fluorimeter. Samples were excited at 295 nm (slit width = 2nm) to avoid excitation of tyrosine. An emission scan was collected from 300 to 400 nm and the ratio of Emission Intensities at 350nm (I350) to 320nm (I320) was plotted against temperature. In the case of lysozyme, the data was then fit to a sigmoidal distribution using the Boltzmann function and Tm was obtained from x$_{o}$, the inflection point of the function,

\begin{equation}
    y = \frac{A_{1}-A_{2}}{1 + e^{(x-x_{0})/dx}} + A_{2}
\end{equation}

For $\beta$-Lactoglobulin, incomplete convergence of the I350/I320 plot over the experimentally feasible temperature range lead to poor sigmoidal fits. As such, thermal stability was assessed by T$_{threshold}$, the temperature at which a pre-determined extent of unfolding, as captured by the ratio of I350/I320 was reached. T$_{threshold}$ was chosen based on the midpoint of the thermal transition in the absence of excipient.

\subsection{Simulated systems}
Computational analyses were performed on the same systems studied by experiment.
Crystallographic structures of Hen egg white lysozyme (PDB ID: 2YVB) and bovine $\beta$-lactoglobulin (PDB ID: 3BLG) were retrieved from the Protein Data Bank. \cite{berman_protein_2000}
In addition, five excipient molecules were considered, i.e. trehalose (PubChem ID: 7427), sucrose (PubChem ID: 5988), sorbitol (PubChem ID 5780), glycerol (PubChem ID: 753) and urea (PubChem ID: 1176). \cite{pubchem_pubchem_nodate}
Starting systems were built with one protein in an osmolyte-water environment, with osmolyte concentrations at either $0\%$, $4\%$, $17\%$, or $37\%$ by weight. 

%

To fix concentration we constructed a dodecahedral box so that each atom of the protein structures was at a distance of at least $1.1\mathrm{nm}$ from the nearest box face. 
We filled one copy of the box with only water molecules and another copy with only molecules of a chemical compound ; we call $w$ the ratio between the number of water molecules ($WM$) and the number of chemical compound molecules ($CM$), that is $r = WM / CM$. 
Since the molecular masses of water ($m$) and chemical compounds ($M$) are known, we have that the concentration is

\begin{equation}
w = \frac{N\cdot M}{N \cdot M + n \cdot m}
\label{eq.1}
\end{equation}

where $w$ is the concentration (known), $N$ is the final number of chemical compound molecules and $n$ is the final number of water molecules.

We can write a system of two equations with two unknowns ($N$ and $n$), where the first equation is \ref{eq.1}, while the second equation is obtained by considering that in the absence of a chemical compound we have $WM$ water molecules and that each molecule replaces $r$ water molecules, therefore

\begin{equation}
\begin{cases}
\displaystyle N = \frac{m}{M} \cdot \frac{w}{1-w}\cdot n \\
\\
\displaystyle N + n = t - rN 
\end{cases}
\end{equation}

\subsection{Molecular dynamics simulations}
All simulations were performed using GROMACS. \cite{van_der_spoel_gromacs_2005}
Topologies of the system were built using the CHARMM-27 force field \cite{brooks_charmm_2009}.
The protein was placed in a dodecahedral simulative box, with periodic boundary conditions, filled with TIP3P water molecules. \cite{jorgensen_comparison_1983}
The 2YVB and 3BLG systems were neutralised with 8 CL atoms and 9 NA atoms respectively.
We chose the CHARMM27 force field and TIP3P water model; we obtained the force field parameters of the chemical compounds from SwissParam. \cite{zoete_swissparam_2011}
Each dynamic extends for $100\mathrm{ns}$ (see next section for details).

The number of molecules used for each simulation are given in Table \ref{sim-details}. 
For all simulated systems, we checked that each atom of the proteins was at least at a distance of $1.1 \, \mathrm{nm}$ from the box borders. 
Each system was then minimised with the steepest descent algorithm. 
Next, a relaxation of water molecules and thermalisation of the system was run in NVT and NPT environments each for $0.1 \, \mathrm{ns}$ at $2 \, \mathrm{fs}$ time-step. 
The temperature was kept constant at $300 \, \mathrm{K}$ with v-rescale thermostat \cite{bussi_canonical_2007}; the final pressure was fixed at $1 \, \mathrm{bar}$ with the Parrinello-Rahman barostat \cite{parrinello_crystal_1980}.

LINCS algorithm \cite{hess_lincs_1997} was used to constraint bonds involving hydrogen atoms.
A cut-off of 12 {\AA} was imposed for the evaluation of short-range non-bonded interactions and the Particle Mesh Ewald method \cite{cheatham_molecular_1995} for the long-range electrostatic interactions.
The described procedure was used for all the performed simulations. 

\begin{table}[]
\centering
\caption[MD simulation details]{MD simulation details. N$_{OS}$ = Total number of osmolyte molecules. N$_{WATER}$ = Total number of water molecules.}
\label{sim-details}
\resizebox{0.4\textwidth}{!}{%
\begin{tabular}{cccccc}
\hline
                                                                            &                            & \multicolumn{2}{c}{2YVB} & \multicolumn{2}{c}{3BLG} \\ \hline
\begin{tabular}[c]{@{}c@{}}Osmolyte \\ Concentration \\ (wt\%)\end{tabular} & Osmolyte                   & N$_{OS}$  & N$_{WATER}$  & N$_{OS}$  & N$_{WATER}$  \\ \hline
4                                                                           & glycerol  & 47.8      & 4836.4       & 41.7      & 3425.8       \\
17                                                                          &                            & 184.6     & 4274.7       & 176.2     & 2987.4       \\
37                                                                          &                            & 414.1     & 3341         & 386.4     & 2343.9       \\ \hline
4                                                                           & sorbitol  & 27.3      & 4843.2       & 22.7      & 3427.8       \\
17                                                                          &                            & 108.2     & 4289.6       & 95.7      & 3007         \\
37                                                                          &                            & 251.6     & 3308.5       & 202.1     & 2440.2       \\ \hline
4                                                                           & sucrose   & 18.3      & 4815.1       & 13.6      & 3402         \\
17                                                                          &                            & 62        & 4269         & 55.1      & 2975.3       \\
37                                                                          &                            & 120.9     & 3626.1       & 108.3     & 2489.6       \\ \hline
4                                                                           & trehalose & 17.9      & 4819.9       & 13.4      & 3406.7       \\
17                                                                          &                            & 55.9      & 4363.6       & 50.7      & 3053.3       \\
37                                                                          &                            & 100.5     & 3827.3       & 112.2     & 2449.9       \\ \hline
4                                                                           & urea      & 76        & 4881         & 65.7      & 3438.1       \\
17                                                                          &                            & 308.3     & 4384         & 289       & 3057.3       \\
37                                                                          &                            & 721.5     & 3452.6       & 659.6     & 2428.5       \\ \hline
\end{tabular}%
}
\end{table}

\subsection*{Interaction energy calculation}

Intra-molecular interaction energies were computed using the parameters obtained from CHARMM force-field. In particular, given two atoms $a_l$ and $a_m$ holding partial charges $q_l$ and $q_m$ , the Coulombic interaction between them can be computed as:
\begin{equation}
\small
\label{eq:C}
E_{lm}^C= \frac{1}{4\pi\epsilon_0}\frac{q_lq_m}{r_{lm}}
\end{equation}
where $r_{lm}$ is the distance between the two atoms, and $\epsilon_0$ is the vacuum permittivity. Van der Waals interactions can instead be calculated as a 12-6 Lennard-Jones potential:
\begin{equation}
\small
\label{eq:LJ}
E_{lm}^{LJ} = \sqrt{\epsilon_l \epsilon_m}\left[ \left(\frac{R_{min}^l + R_{min}^m}{r_{lm}}\right)^{12} - 2\left(\frac{R_{min}^l + R_{min}^m}{r_{lm}}\right)^{6}\right]
\end{equation}
where $\epsilon_l$ and $\epsilon_m$ are the depths of the potential wells of $a_l$ and $a_m$ respectively, $R_{min}^l$  and $R_{min}^m$ are the distances at which the potentials reach their minima.

The total interaction energy  between each couple of residues is defined as:

\begin{equation}
    E_{AA_{ij}}^X= \sum_{l = 1}^{N_{atom}^i }\sum_{m = 1}^{N_{atom}^j} E_{lm}^X
\end{equation}

where $E_{AA_{ij}}^X$ is the energy between two amino acids $i$ and $j$, obtained as the sum of the interactions between each atom of the two residues ($N_{atom}^i$, $N_{atom}^j$); $X$ stands for the kind of interaction considered, either Coulombic ($X = C$) or Lennard-Jones ($X = LJ$).

As for the distance between a pair of residues, this was assessed by selecting the minimum distance between the atoms composing them.

\subsection{Hydrogen bond calculation}
For each of the dynamics we extracted one frame every 1 ns for a total of 101 frames. For each frame, using the Chimera software, we counted the number of hydrogen bonds that each residue forms with other residues, with water molecules and with chemical compounds.

\subsection{Calculation of radial distribution functions}
 Radial distribution functions (g(r)) for protein-osmolyte and protein-water interactions were calculated by the equation below:
 
 \begin{equation}
     g(r)_{a,b} = \frac{1}{\rho_{bulk,b}} \frac{\delta N_{b}(r)}{\delta V(r)} 
 \end{equation}
 wherein $\delta$V(r) and $\delta$N$_{b}$(r) indicate the volume and the number of b particles corresponding to the bin defined by (r, dr), respectively; r is the distance from reference particle a and dr is the bin size. 
 [Ma, Cui 2003]

\subsection{Calculation of preferential interaction parameters}
Osmolyte-protein preferential interaction parameters ($\Gamma_{os. , prot}$) were calculated using the equation, 

\begin{equation}
    \Gamma_{os., prot.} = \langle n^{local}_{os.} - \frac{n^{bulk}_{os.}}{n^{bulk}_{water}} (n^{local}_{water}) \rangle
\end{equation}

 wherein $\langle$ $\rangle$ indicates the ensemble average, $n^{local}_{os.}$ and $n^{bulk}_{os.}$ are the number of osmolyte molecules in the local and bulk environments, respectively, and  $n^{local}_{water}$ and $n^{bulk}_{water}$ are the number of water molecules in the local and bulk environments. \cite{shukla_molecular_2009} The cutoff distance marking the boundary between bulk and local domains, r*, was set to the distance at which bulk concentration reached equilibrium; this was found to be 15.0{\AA} for LYS and 17.6{\AA} for $\beta$L by examination of the protein-osmolyte and protein-water radial distribution functions. The number of osmolyte or water molecules was calculated via integration of the respective protein-osmolyte or protein-water radial distribution function:
 
 \begin{equation}
n_{(r_{a}, r_{b})} = 4\pi  * \rho * \int_{r_{a}}^{r_{b}} g(r) r^2 dr 
\end{equation}



\subsection{Graph analysis}
Perturbations to protein structure and osmolyte clustering behaviour were studied by graph analysis. Protein intermolecular bonds were modelled as residue interaction networks (RINs) wherein residue C$\alpha$ atoms were approximated as nodes and two nodes were considered to be in contact if their distance was less than 9{\AA}. The weight for each link was determined by interaction energy (van der Waals or Coulombic). For each RIN, average metrics were calculated considering only key residues selected based the threshold criteria ($E < -10$, degree $>25$). Calculations were performed on a total of 10 trajectories for each discrete simulation.

Osmolyte-osmolyte clustering was studied via graph analysis. 
Osmolyte systems were modelled as unweighted graphs using NetworkX package in Python. \cite{hagberg_exploring_2008}
Discrete molecules were represented as nodes by their centre of mass.
A link existed between two nodes when the distance between them was less than the average H-bond osmolyte-osmolyte interaction distance defined by g(r) (2.65{\AA}).
To account for molecular (node) size, this value was corrected by 2 * R$_{g}$ for each osmolyte.
Cliques were defined as complete subgraphs wherein a link existed between all possible node pairs.
Average clique size was calculated by NetworkX and volume-normalised using RDKit generated molecular volumes for each species. \cite{noauthor_rdkit_nodate}
All metrics were calculated as the time-averages over 101 frames extracted every 1ns of simulation time.

Normalised betweenness centrality metrics for residue interaction and osmolyte networks were calculated according to the equation, 
\begin{equation}
    BC(v) = \frac{\sum_{s\neq v \neq t} \frac{\sigma_{st}(v)}{\sigma_{st}}}{(n-1)(n-2)/2}
\end{equation}
wherein the $BC(v)$ is the betweenness centrality of node $v$, $n$ is the number of nodes in the graph, $(s,t)$ represents each pair of vertices, $\sigma_{st}$ is the total number of shortest paths from node s to node t and  $\sigma_{st}(v)$ is the total number of shortest paths that pass through the non-terminal node $v$. 

\section*{Competing Interests}
The authors declare no conflict of interest.

\section*{Acknowledgements}

M.M., G.R., and G.G.T. acknowledge support by European Research Council through its Synergy grant programme, project ASTRA (grant agreement No 855923) and by European Innovation Council through its Pathfinder Open Programme, project ivBM-4PAP (grant agreement No 101098989).
N.W. and O.A.S acknowledge support by AB Agri grant number RG85712.

\section*{Author contributions}

M.M., N.W., O.A.S. and E.M. conceived and devised research. G.R. and G.G.T. contributed additional ideas.   N.W. performed experiments and analysed data. M.M carried out numerical simulations and statistical analyses. M.M. and N.W. wrote the manuscript. All authors revised the manuscript.


\newpage

\section{Supplementary Information}

\newpage

\subsection{Raw RMSF Data}

\begin{figure}[]
    \centering
    \includegraphics[width = 0.8 \textwidth]{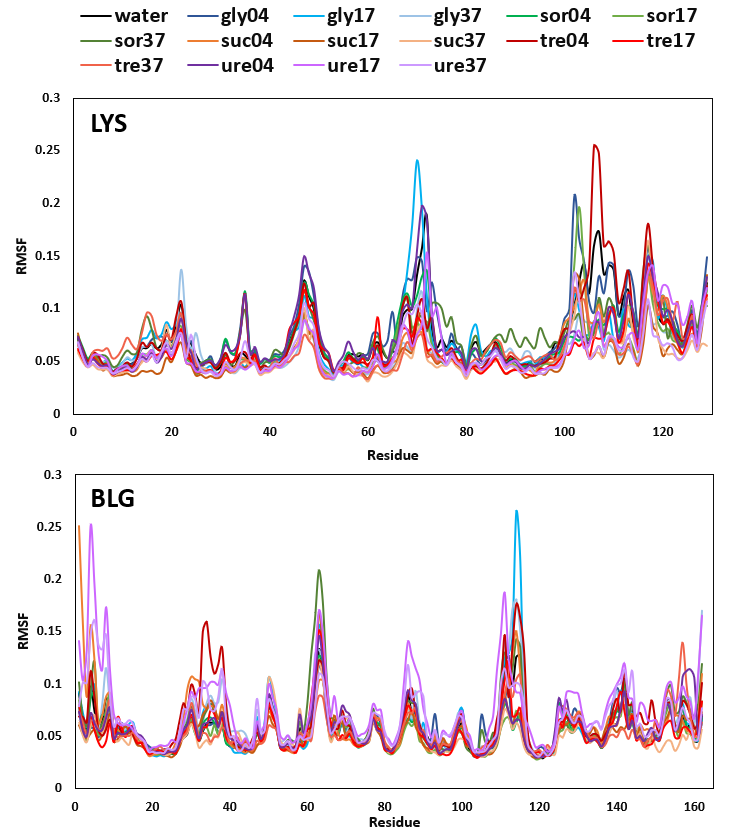}
    \caption{Raw RMSF data for BLG and LYS across all systems.}
    \label{raw-rmsf}
\end{figure}

\subsection{Osmolyte-Protein Radial Distribution Functions}

\subsection{Water-Protein Radial Distribution Functions}

\begin{figure}[]
    \centering
    \includegraphics[width = \textwidth]{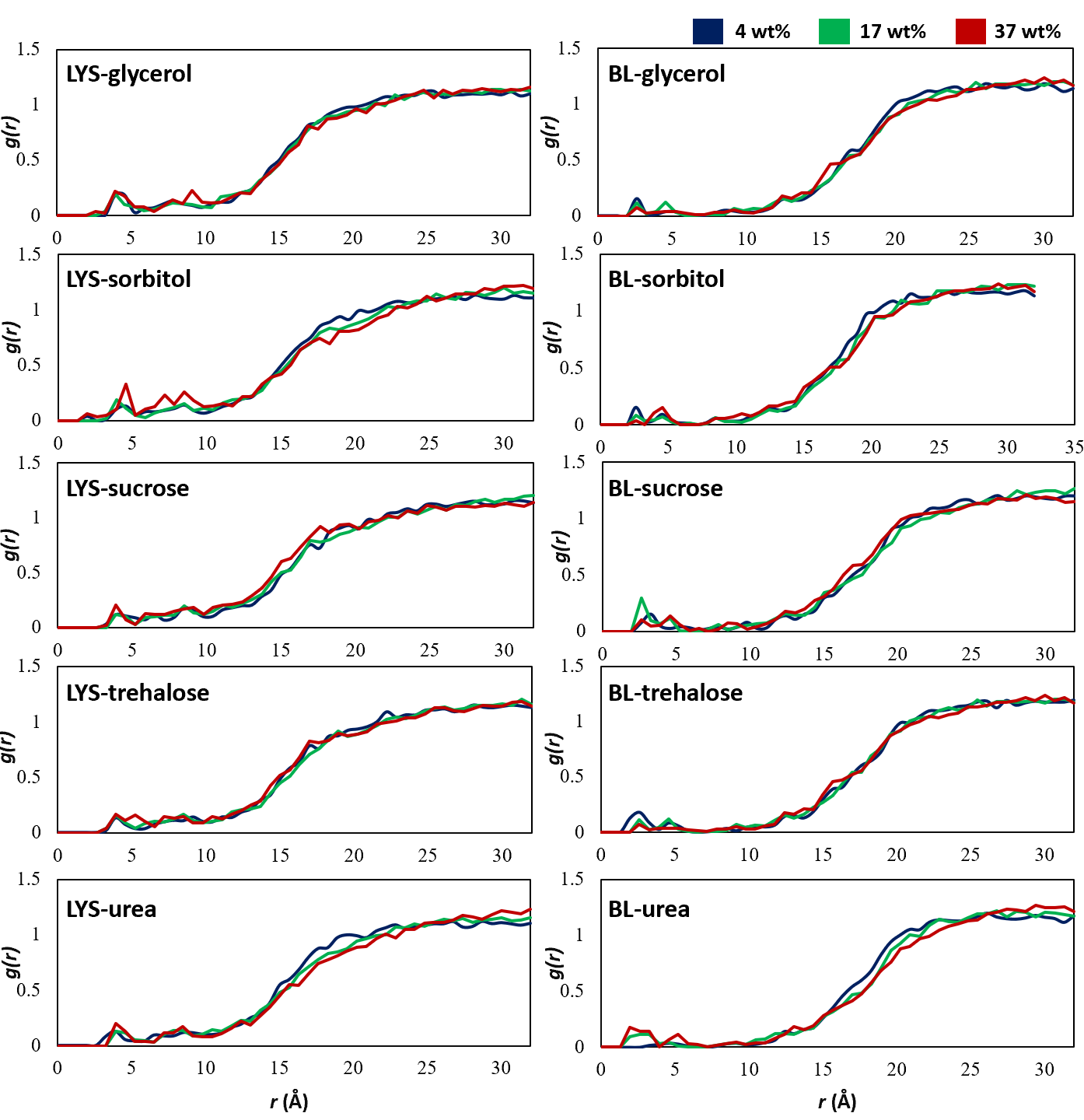}
    \caption{Water-protein radial distribution functions.}
    \label{water-protein rdf}
\end{figure}

\subsection{Room Temperature ITF Measurements}

\begin{figure}[]
    \centering
    \includegraphics[width = \textwidth]{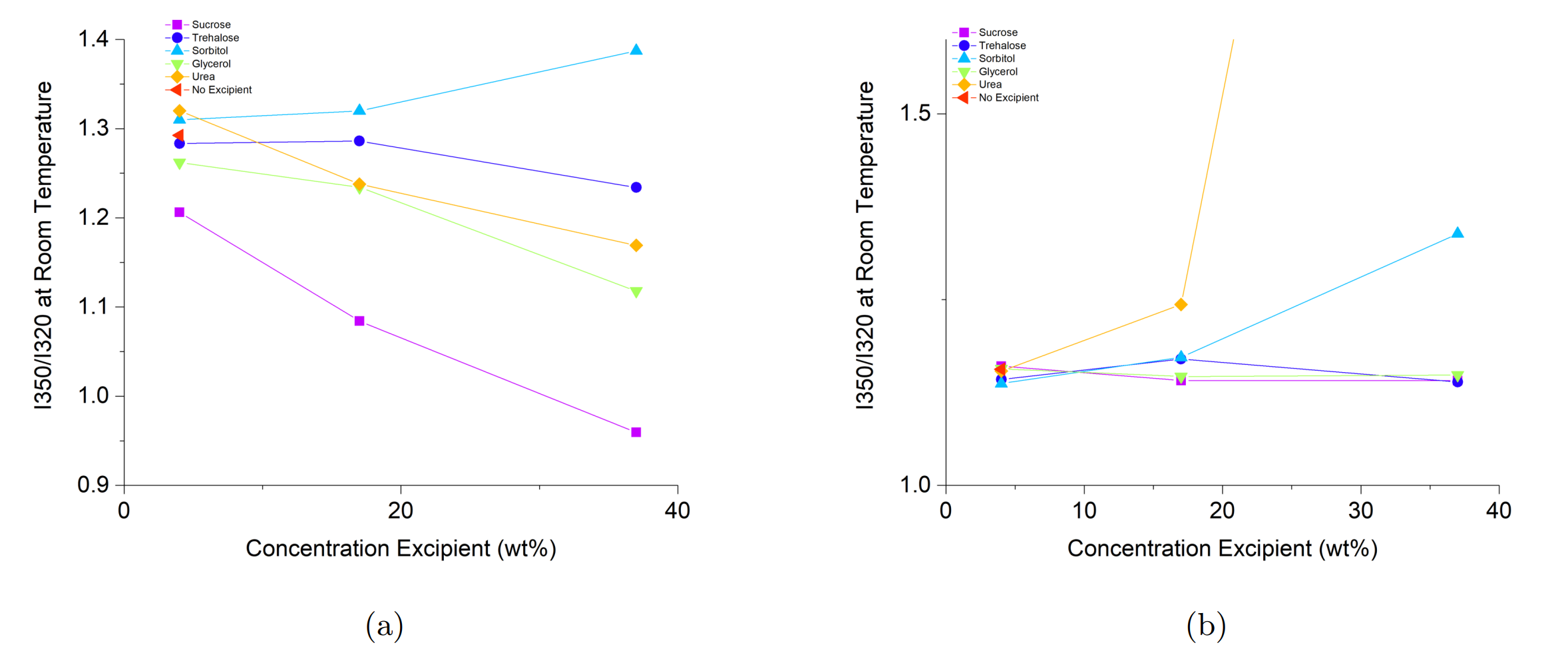}
\caption{Excipient-induced room-temperature ITF shifts for \textbf{a)} LYS and \textbf{b)} $\beta$L. \label{rt-itf}}
\end{figure}

\subsection{MD Snapshots}
\begin{figure}[]
    \centering
    \includegraphics[width=0.8\textwidth]{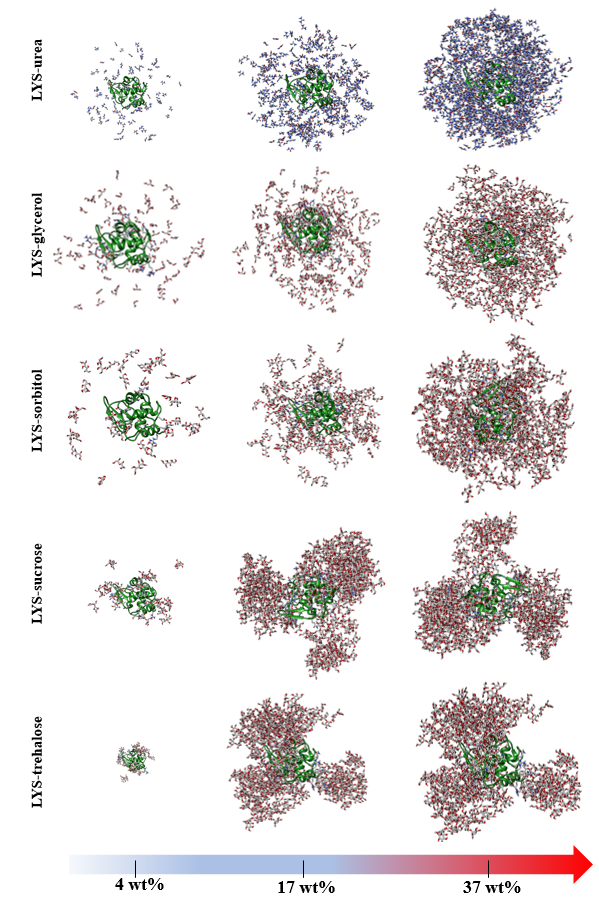}
    \caption{LYS MD Snapshots}
    \label{snapshots-LYS}
\end{figure}

\begin{figure}[]
    \centering
    \includegraphics[width=0.8\textwidth]{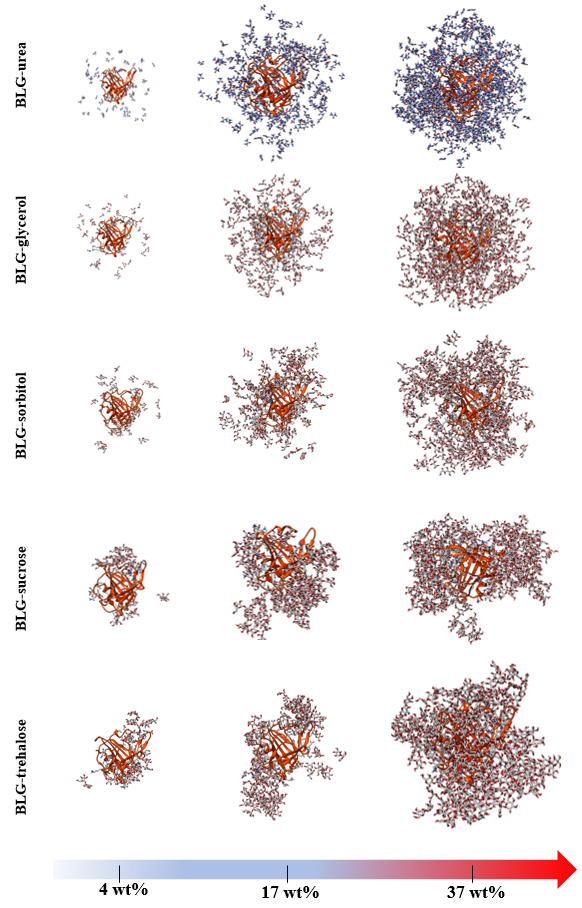}
    \caption{BLG MD Snapshots}
    \label{snapshots-BLG}
\end{figure}

\end{document}